# Enhancement of the electromechanical response in ferroelectric ceramics by design

K. P. Jayachandran,[a)] J. M. Guedes, and H. C. Rodrigues

*IDMEC-IST, Department of Mechanical Engineering, Technical University of Lisbon, Av. Rovisco Pais, 1049-001 Lisbon, Portugal*

It is demonstrated based on continuum mechanics modeling and simulation that it is possible to obtain polycrystalline ceramic ferroelectric materials which beggars single crystals in electromechanical properties. The local inhomogeneities at the ferroelectric domain-scale level due to spontaneous polarization and the underlying anisotropy are taken into consideration in the framework of mathematical homogenization of physical properties in ferroelectric materials. The intrinsic randomness of the spatial distribution of polarization is shown to be judiciously employed for the design of better polycrystalline ferroelectrics. The noncollinear rotation of the net polarization-vectors embedded in crystallites of the ceramic ferroelectrics is demonstrated to play the key role in the enhancement of physical properties.

---

[a)] Electronic mail: jaya@dem.ist.utl.pt

## I. INTRODUCTION

The phase transition from a centrosymmetric paraelectric phase to a lower symmetric ferroelectric phase results in domains possessing spontaneous polarizations juxtaposed but separated by walls.[1, 2] It is demonstrated both experimentally [3-6] and theoretically [7-10] that a number of ferroelectric (FE) single crystals including $BaTiO_3$ exhibits an enhanced piezoelectric strain when poled along a nonpolar axis. This orientation-dependent enhancement of piezoelectricity is mainly attributed to noncollinear polarization rotation. [3, 10-12] The noncollinearity stems essentially from the domain structure of the single crystals. Recent reports [13-15] on ceramic $BaTiO_3$ suggest that grain-orientation in a flux of random grain boundaries could greatly enhance the piezoelectricity. These findings underscore previous notion [16] of the role of grain boundaries in the control and design of ceramic FE materials. Yet, studies on polarization rotation are limited in ceramics owing to many reasons though ceramics present ease in manufacture and in compositional modifications and represent the widest application area of FE materials. This impedes the advance and growth of new high performance ceramic materials which are simultaneously cost-effective and reproducible that beggars the single crystals -which are prone to deficiencies such as depolarization and chemical inhomogeneity- in piezoelectric performance. Computationally intensive atomistic models have size limitations in ceramics as the number of constituent atoms in a statistical sample of it is prohibitively high so that an exact calculation in the state space is no longer possible. But continuum models incorporating domain-scale information like polarization-vector orientation and distribution have been used to describe the macroscopic features of ferroelectrics. [7, 8, 17] Nevertheless, the domain pattern observed in FE polycrystals is however not unique, but depends on many factors like the composition, the loading pattern, intergranular stresses developed during production, uncompensated surface charges and physical imperfections. [18, 19] Thus a model based on a certain domain pattern as used in micromechanics averaging methods [7, 8, 20] will be highly discriminatory and limited in its representation of a real FE polycrystal. In this context continuum models which accommodates the net polarization and the resulting electromechanical response of each constituent crystallites (grains) of a statistical sample of FE ceramic present a plausible alternative.

In this paper we propose a design strategy based on a continuum mechanics and modeling to attain electromechanic figures of merit in FE ceramics and single crystals. We apply this methodology to show the piezoelectric enhancement in classic perovskite ceramic $BaTiO_3$ and demonstrate its supremacy over single crystal $BaTiO_3$. Also the single crystal characteristics as a function of orientation are analyzed comprehensively to have a better knowledge of polycrystal response. Recently we have shown that orientation distribution of polarization in a ceramic ferroelectric material is critical in its piezoelectric efficiency. [21] The macroscopic properties of a ceramic FE in general differ significantly from those of single crystals mainly due to the imperfect alignment of the crystallographic axes of the constituent crystallites. The as-grown, virgin FE polycrystalline structure results in a near complete compensation of polarization, if any, until coaxed with a large electric field (poling). [16] Generally, for many ferroelectrics, the polarization-vector $\boldsymbol{P}$ rotates with an electric field applied along a nonpolar direction [10] and the polarization rotation in such cases does not necessarily evolve though a monoclinic phase. [20] A crystallite of a ceramic FE possesses a net polarization $\bar{\boldsymbol{P}}$ due to the additive contributions of spontaneous polarizations $\boldsymbol{P}_s$ of various FE domains evolved. In an aggregate of crystallites like in FE ceramics the net polarization $\bar{P}$ of each crystallite has unique orientation in space. Thus $\bar{P}(\equiv \bar{P}(\phi,\theta,\psi))$ require three Euler angles ($\phi,\theta,\psi$) to describe its (and the crystallite's) orientation with reference to a fixed macroscopic coordinate system **x**. In fact, $\bar{P}$ manifests in a crystallographic coordinate system embedded in the crystallite. The resolved vector components of $\bar{P}$ are statistically distributed in a uniform distribution in as-grown unpoled FE ceramic material. Hence the angles ($\phi,\theta,\psi$) too follow a uniform distribution pattern. However, after poling the $\bar{P}$ s would be nearly aligned with the electric field (keeping in mind that some crystallites will not have possible domain states that are nearly aligned with the field).[22] The crystallite orientations would follow a Gaussian pattern with a probability distribution function (PDF) $f(\alpha|m,\sigma) = (1/\sigma\sqrt{2\pi})\exp-\left[(\alpha-m)^2/2\sigma^2\right]$.[21] Here $\alpha$ is the random variable which are the orientations represented by the Euler angles $\theta$, $\phi$ and $\psi$ and $m$ and $\sigma$ are the parameters of the distribution viz., the mean and the standard deviation respectively.

## II. THEORETICAL AND COMPUTATIONAL METHODS

In FE ceramics the electrical and mechanical fields are spatially distributed and nontrivially coupled through the direct or converse piezoelectric effects, the local spatial compatibility of the domains, and the anisotropy of the underlying dielectric and elastic properties. Hence the electrical and elastic boundary conditions and the orientation of the polarization and permittivity axes of crystallites complicate the microscopic analysis of the polycrystal. We have used the mathematical homogenization method [23, 24] which efficiently characterizes the equilibrium macroscopic electromechanical properties of FE. The mathematical theory of the homogenization method accommodates the interaction of different phases in characterizing both the macro- and the micro-mechanical behaviors. In homogenization theory the material is locally formed by the spatial repetition of very small unit-cells, when compared with the overall macroscopic dimensions. Further, the material properties are periodic functions of the microscopic variable, where the period is very small compared with the macroscopic variable. This enables the computation of equivalent material properties by a limiting process wherein the microscopic cell size is approaching zero.

In a piezoelectric medium the mechanical stress $T_\mu$ and the electrical displacement $D_i$ relate to the mechanical strain $\varepsilon_\nu$ and electrical field intensity $E_k$ by the constitutive relations: $T_\mu = C^E_{\mu\nu}\varepsilon_\nu - e_{k\mu}E_k$ and $D_i = e_{i\nu}\varepsilon_\nu + \kappa^\varepsilon_{ik}E_k$. Here the Greek indices running from 1 to 6 abbreviates the tensorial forms represented by the Latin indices running from 1 to 3. $C^E_{\mu\nu}$ and $e_{k\mu}$ are the stiffnesses under short-circuit boundary conditions and the piezoelectric stress coefficients respectively and $\kappa^\varepsilon_{ik}$ are the free-body dielectric permittivity. Symmetry requires that $\kappa^\varepsilon_{ij} = \kappa^\varepsilon_{ji}$, $e_{kij} = e_{kji}$ and $C^E_{\mu\nu} = C^E_{\nu\mu}$. The microstructure or unit-cell (or representative volume element) used in our simulation is made up of randomly oriented grains or crystallites. Orientation of the crystallite is governed by the orientation of the net polarization-vector $\bar{P}$. The macroscopic piezoelectric material is constructed by the contiguous assembly of a large number of identical representative unit-cells.

The asymptotic analysis and homogenization of the piezoelectric medium (cf. the appendix) has resulted in the macroscopic piezoelectric

$$e^H_{prs}(\boldsymbol{x}) = \frac{1}{|Y|}\left\{\int_Y \left[e_{kij}(\boldsymbol{x},\boldsymbol{y})\left(\delta_{kp} + \frac{\partial R^{(p)}}{\partial y_k}\right)\left(\delta_{ir}\delta_{js} + \frac{\partial \chi^{(rs)}_i}{\partial y_j}\right)\right.\right.$$

$$-e_{kij}(\boldsymbol{x},\boldsymbol{y})\frac{\partial \Phi_i^{(p)}}{\partial y_j}\frac{\partial \psi^{(rs)}}{\partial y_k}\Bigg]dY\Bigg\} \quad (1),$$

dielectric

$$\kappa_{pq}^{\varepsilon\ H}(\boldsymbol{x}) = \frac{1}{|Y|}\Bigg\{\int_Y\Bigg[\kappa_{ij}^{\varepsilon}(\boldsymbol{x},\boldsymbol{y})\Bigg(\delta_{ip}+\frac{\partial R^{(p)}}{\partial y_i}\Bigg)\Bigg(\delta_{jq}+\frac{\partial R^{(q)}}{\partial y_j}\Bigg)$$

$$-e_{kij}(\boldsymbol{x},\boldsymbol{y})\Bigg(\delta_{kp}+\frac{\partial R^{(p)}}{\partial y_k}\Bigg)\frac{\partial \Phi_i^{(q)}}{\partial y_j}\Bigg]dY\Bigg\} \quad (2)$$

and elastic

$$C_{rspq}^{E\ H}(\boldsymbol{x}) = \frac{1}{|Y|}\Bigg\{\int_Y\Bigg[C_{ijkl}^{E}(\boldsymbol{x},\boldsymbol{y})\Bigg(\delta_{ip}\delta_{jq}+\frac{\partial \chi_i^{(pq)}}{\partial y_j}\Bigg)\Bigg(\delta_{kr}\delta_{ls}+\frac{\partial \chi_k^{(rs)}}{\partial y_l}\Bigg)$$

$$+e_{kij}(\boldsymbol{x},\boldsymbol{y})\Bigg(\delta_{ip}\delta_{jq}+\frac{\partial \chi_i^{(pq)}}{\partial y_j}\Bigg)\frac{\partial \psi^{(rs)}}{\partial y_k}\Bigg]dY\Bigg\} \quad (3)$$

coefficients in index notation. $\chi_i^{(rs)}$, $R^{(p)}$, $\Phi_i^{(p)}$ and $\psi^{(rs)}$ are characteristic functions of the unit-cell of size $Y$ and $\delta$ is the Kronecker delta symbol. The functions $C_{ijkl}^E$, $\kappa_{ij}^\varepsilon$ and $e_{kij}$ can be described in microscopic coordinate system $y_i$ (which coincides with the macroscopic frame **x**) using the components of Euler transform tensors [25] from crystallographic coordinates. The superscript $H$ would be dropped henceforth from the homogenized coefficients for brevity. The elastic compliance $s_{\mu\nu}$ is the reciprocal of the stiffness $C_{\mu\nu}$ and the piezoelectric strain coefficients are $d_{i\nu} = \sum_{\mu=1}^{6}e_{i\mu}s_{\mu\nu}$.

We developed a three-dimensional (3D) numerical model to compute the homogenized piezoelectric properties of ferroelectric ceramics. Further, the numerical solution of the coupled piezoelectric problem is sought using the finite element method to eventually compute the homogenized coefficients. As an important application of this model, we compute the effective macroscopic electromechanical properties as a function of the rotation of the tetragonal (*T*) ferroelectric BaTiO$_3$. First we make the microscopic coordinates $y_i$ in the space of $\mathbb{R}^3$ and the crystallographic coordinate system *xyz* coincides. Then the Euler angles $(\phi, \theta, \psi)$ are defined in the following way;[25] firstly, the crystal is rotated by angle $\phi$ around the *z*-axis, then rotate an angle $\theta$ around the new *x*-axis, and finally by an angle $\psi$ around the new *z*-axis. All rotations are in the positive direction (counter-clockwise). A comprehensive survey of the entire piezoelectric anisotropy of the crystal is possible with the variation of $(\phi, \theta, \psi)$ through its full range. As standard deviation $\sigma \to 0$ one can approach perfect texture similar to single crystals and $\sigma \to \infty$ generates a randomly oriented (net polarizations $\bar{P}$) polycrystal with no aggregate piezoelectricity. The texture parameters of PDF, $\sigma$ and *m* are obviously related to the material processing conditions like poling electric field and/or annealing temperature. A material is termed textured if the crystallites are aligned in a preferred orientation along certain lattice planes. One can view the textured state of a material (typically in the form of thin films) as an intermediate state in between a randomly oriented polycrystalline powder and a fully oriented single crystal.

Since our model is able to predict the macroscopic electromechanical coefficients of all piezoelectric crystal classes, application of the same to ferroelectric BaTiO$_3$ requires the knowledge of symmetry which is tetragonal P*4mm*. The polycrystal sampling is performed using a unit-cell of $20 \times 20 \times 20$ mesh with 64000 integration points and each node of the trilinear solid element is allowed four degrees of freedom (one electric potential and three displacements). Full integration (2-point Gaussian integration rule in each direction) is used for the evaluation of the stiffness, piezoelectric and dielectric matrices and for the homogenization. As the unit-cell is expected to capture the response of the entire piezoelectric system (where the macroscopic piezoelectric material is constructed by adding, contiguously to the representative unit-cell, a large number of other identical unit-cells), particular care is taken to ensure that the deformation across the boundaries of the representative unit-cell are compatible with the deformation of adjacent unit-cells. Hence all the load cases are solved by enforcing periodic boundary conditions in the unit-cell for the displacements and electrical potentials. Each finite element is assigned to represent a single grain of the polycrystal. The numerical simulation of ceramic $BaTiO_3$ are done using the parameters of single crystal data from Ref. [26] using the present homogenization model computationally implemented in FORTRAN. We have used a random number generator to generate the orientation ($\theta$) distribution of grains in the polycrystal. An important aspect here is that for the same value of the parameters $\sigma$ and $m$ the generator delivers different sets of data (angles) each time it is invoked. Thus to verify the robustness of the proposed model, we generated different sets of random orientations, for a specific value of the parameter $\sigma$ and $m$, and performed the respective simulations varying the number of crystallites in the microstructure (i.e., varying the number of finite elements used to discretize the unit-cell) to check the possibility of scatter of homogenized results. This in turn becomes another experiment to arrive at an optimal unit-cell that can statistically represent the microstructural features of the polycrystal.[27] The convergence of electromechanic properties with unit-cell size pave the way to determine the simulation-space-independent, macroscopic properties.

## III. RESULTS AND DISCUSSION

**A. Polarization rotation and the single crystal**

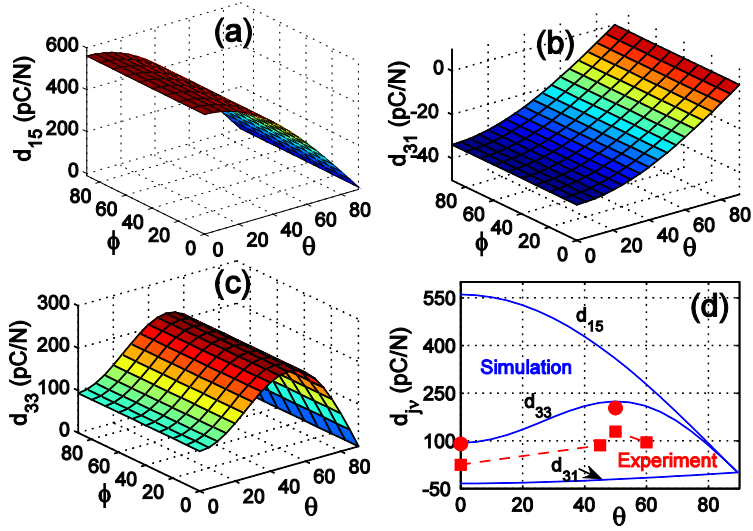

FIG.1. Variation of the piezoelectric strain coefficients (a) $d_{15}$, (b) $d_{31}$ and (c) $d_{33}$ with the orientation determined by Euler angles $\phi$, $\theta$ (in degrees) in BaTiO$_3$ single crystal. (d) Cross-section curves obtained when a cut by the crystallographic plane (010) to enable comparison of simulated $d_{33}$ with experiments in Ref. [12] (filled circles) and Ref. [29] (filled squares).

In this section we analyse the impact of the crystal orientation on the electromechanical response of single crystal BaTiO$_3$. The main aim of this section is to demonstrate the robustness of the present method compared to available experimental and theoretical results. Here we perform simulation by discretizing the single crystal unit-cell into $20\times 20\times 20$ mesh. The Euler angles $\theta$ and $\phi$ are simultaneously varied with reference to the poling direction [001]. For this study we performed simulation on single crystal model by rotating the angle $\theta$ from 0 to 90° at each orientation of $\phi$, keeping $\psi$ at zero. Then $\phi$ is varied from 0 to 90° to survey the full anisotropy of the BaTiO$_3$ spanning through the entire crystallographic space bound by the panes {001}, {010} and {100}. To check the accuracy of our model we proceed with the comparison of our simulation results in single crystals with the experiment. First, we have compared the $s_{\mu\nu}$ and $d_{i\nu}$ obtained from the simulations in $T$ phase single crystal BaTiO$_3$ oriented through the spontaneous polar axis [i.e., along the [001] direction, where the Euler angle ($\phi, \theta, \psi$) is zero] with the experiment [26] and have found excellent agreement between the two. [21]

Next, we discuss the effect of canting the single crystal. $d_{j\nu}$ are the coefficients connecting the electric field to the strain in converse piezoelectricity. The shear and transverse piezoelectric coefficients $d_{15}$ and $-d_{31}$ shown in Figs. 1(a) and (b) respectively are decreasing monotonously with the crystal orientation. The longitudinal component of piezoelectric tensor ($d_{33}$) describes the coupling of the piezoelectric strain $\varepsilon_3$ to the electric field $E_3$ applied along the direction of the measured strain by $\varepsilon_3 = d_{33} E_3$. $d_{33}$ of BaTiO$_3$ in Fig. 1(c) peaks at $\theta \approx 50°$, and reaches a value $d_{33} = 223.7$ pC/N which is more than twice as large as the original value 94 pC/N of the single crystal poled along [001]. In fact, in varying the angle $\phi$, we aim to make our simulation pass through the vicinity of the critical orientation {111} besides surveying the spatial piezoelectric anisotropy. As is revealed from Fig. 1(c) the rotation $\phi$ of the $ab$-plane has no effect on $d_{j\nu}$. Thus the single crystal canted along {111} also has $d_{33} = 223.7$ pC/N. This result is consistent with the findings from the energy minimization techniques [8] and phenomenological studies [9] of BaTiO$_3$. The present $d_{33}$ values compares well with the measurements of Wada *et. al.* [12] for [001]-oriented ($d_{33}$ = 90 pC/N) and [111]-oriented ($d_{33}$ = 203 pC/N) engineered-domain BaTiO$_3$ [Fig.1 (d)]. [28] A similar piezoelectric anisotropy is experimentally confirmed by Du *et. al.* [29] in partially depoled single crystalline BaTiO$_3$ where the $d_{33}$ is shown to peak at $\theta = 50°$ with a value 129.44 pC/N from 24.69 pC/N at [001]-orientation. Nevertheless, they have obtained a theoretical $d_{33}$ of 250 pC/N for the canted BaTiO$_3$ along {111} in the same study. This is an interesting feature since in materials possessing lower symmetries like in the rhombohedral PMN–PT, the enhancement of longitudinal piezoelectric strain occurs at a different orientation. [3, 30] Also all the piezoelectric moduli of BaTiO$_3$ drops to zero at the orientation $\theta = 90°$. The $T$ phase of BaTiO$_3$ has $d_{33} = 2e_{31}s_{13}+e_{33}s_{33}$. The product $e_{33}s_{33}$ is one order of magnitude greater than $2e_{31}s_{13}$. Thus the fluctuation in $e_{33}s_{33}$ will reflect in $d_{33}$ significantly. In our simulation the piezoelectric coefficient $e_{33}$ reaches a value of 27.32 C/m$^2$ between orientations $\theta = 50$ and 55° from a spontaneous polarization value 6.7 C/m$^2$. This four-fold enhancement contributes significantly to the peaking of $d_{33}$ at $\theta \approx 50°$. In addition to this, $e_{33}$ and $s_{33}$ after coordinate transformation becomes a linear combination of $e_{i\nu}$ and $s_{\mu\nu}$ such that

$$e_{33} = (e_{15}^c + e_{31}^c)S^2\theta S^2\phi C\theta + e_{33}^c C^3\theta + (e_{31}^c C\phi - e_{15}^c S\phi)S^2\theta C\theta C\phi$$ and

$$s_{33} = s_{33}^c C^4\theta + s_{11}^c(S^4\theta S^4\phi + S^4\theta C^4\phi) + (2s_{12}^c + s_{66}^c)S^4\theta S^2\phi C^2\phi + (2s_{13}^c + s_{44}^c)S^2\theta C^2\theta$$ (here

the quantities bearing superscript $c$ are expressed in the crystallographic coordinate system and $S$ and $C$ refers to trigonometric *sine* and *cosine*). Hence all nonvanishing components of piezoelectric ($e_{i\nu}$) and compliance ($s_{\mu\nu}$) coefficients play a role in determining $d_{33}$. At $\theta \approx 50°$ the contribution from the first term is about seven times bigger than that from the second term because of the bigger value of the shear modulus $e_{15}^c$ (= 34.2 C/m$^2$) compared to $e_{33}^c$ (= 6.7 C/m$^2$).

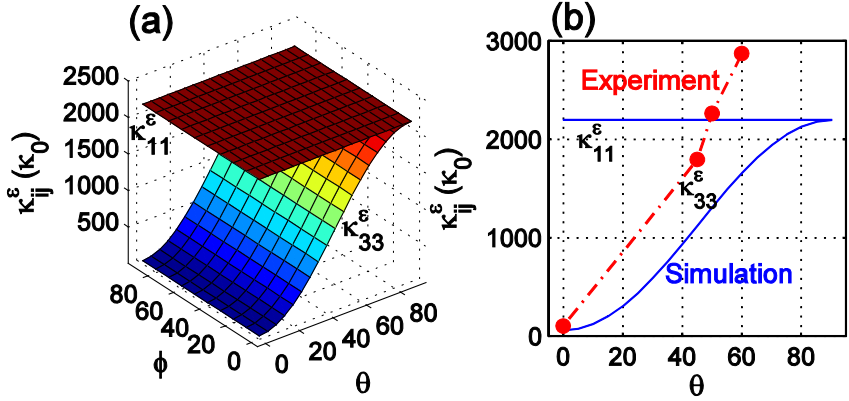

FIG.2. (a) Dependence of the relative dielectric permittivities $\kappa_{ij}^{\varepsilon}$ (in units of $\kappa_0$, the dielectric permittivity of free space) on the crystal orientation determined by Euler angles $\phi$, $\theta$ (in degrees) in BaTiO$_3$ single crystal. (b) Cross-section curves obtained when a cut by the crystallographic plane (010) to enable comparison of simulated $\kappa_{33}^{\varepsilon}$ with experiments in Ref. [29] (filled circles).

The permittivity $\kappa$ is inextricably related to the dielectric susceptibility $\chi$ by $\kappa = \kappa_0(1+\chi)$, where $\kappa_0$ is the permittivity of vacuum.[31] The anisotropy of dielectric susceptibility is critical to phase transition in ferroelectrics.[32] The orientation analyses from Figs. 2(a) and their cross-section along (010) plane shown in the Fig. 2(b) shows large anisotropy of the longitudinal dielectric coefficient $\kappa_{33}^\varepsilon(\kappa_0)$. It increases rapidly with the orientation and reaches the areal component $\kappa_{11}^\varepsilon(\kappa_0)$ value at $\theta = 90°$. The trend shown by $\kappa_{33}^\varepsilon(\kappa_0)$ of BaTiO$_3$ is similar with the reported experimental study,[29] despite exhibiting progressive shits from the beginning [$\kappa_{33}^\varepsilon(\kappa_0)$ = 103 $\kappa_0$ in the [001] direction compared to our value of ($\kappa_{33}^\varepsilon(\kappa_0)$ = 56 $\kappa_0$)]. While $\kappa_{11}^\varepsilon$ is invariant under coordinate transformation, $\kappa_{33}^\varepsilon$ varies according to $\kappa_{33}^\varepsilon = \kappa_{11}^\varepsilon S^2\theta + \kappa_{33}^\varepsilon C^2\theta$. This is clearly reflected in Fig.2 where $\kappa_{11}^\varepsilon(\kappa_0)$ is independent of ($\theta$, $\phi$) and $\kappa_{33}^\varepsilon(\kappa_0)$ varies with $\theta$ and not with $\phi$.

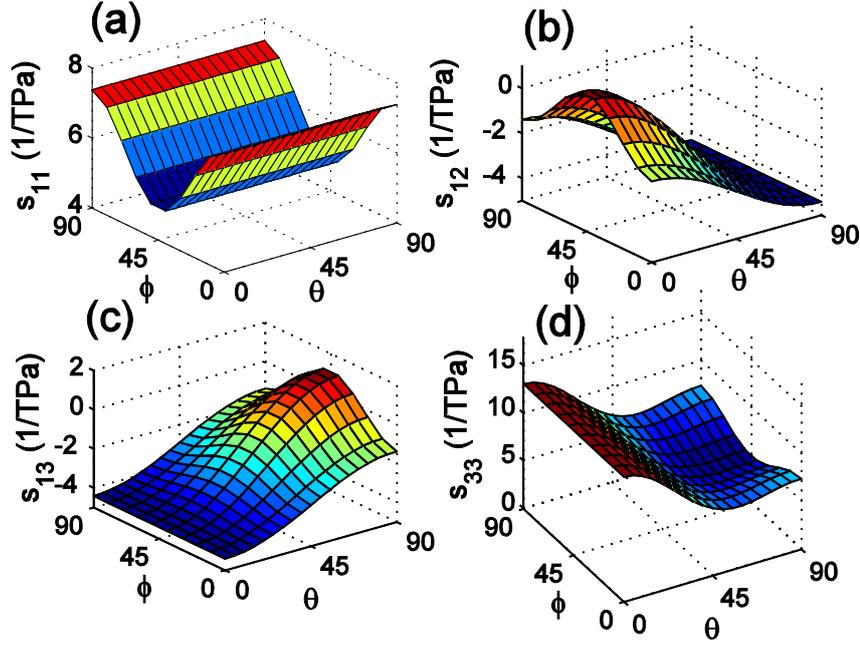

FIG.3. Variation of the elastic compliances (a) $s_{11}$, (b) $s_{12}$, (c) $s_{13}$ and (d) $s_{33}$ with the crystal orientation $\phi, \theta$ (in degrees) in BaTiO$_3$ single crystal.

The longitudinal compliance coefficient $s_{11}$ characterising the stiffness along the *ab*-plane exhibits strong anisotropy with $\phi$ while remaining independent of the orientation $\theta$. The value of $s_{11}$ falls from 7.38 (TPa)$^{-1}$ at $\phi = \theta = 0$ (i.e. along [001] poling direction) to 4.94 (TPa)$^{-1}$ at $\phi = \theta = 50°$ [Fig. 3(a)]. Unlike piezoelectric strains $d_{jv}$ and permittivities $\kappa^{\varepsilon}_{ij}$ of *T* phase BaTiO$_3$ as seen above, $s_{11}$ depends only on $\phi$ and not on $\theta$ but the rest of $s_{\mu\nu}$ do vary with both the angles. The off-diagonal compliances $s_{12}$ shown in Fig. 3(b) and $s_{13}$ shown in Fig. 3(c) exhibit dependence on both $\phi$ and $\theta$. The $s_{12}$ of BaTiO$_3$ increases almost two-fold to reach 1.06 (TPa)$^{-1}$ at about $\phi = 50°$, and $\theta = 0$ from the [001]-poled value of -1.39 (TPa)$^{-1}$. However, it decreases with $\theta$, the rotation of the *c*-axis of the crystal from the poling field direction. Interestingly, the value of the compliance $s_{13}$ also attains 1.06 (TPa)$^{-1}$ at about $\phi = 50°$ however at $\theta =$

90°. The $s_{13}$ of a [001]-oriented single crystal BaTiO$_3$ is -4.41 (TPa)$^{-1}$. The dependence of $s_{33}$ [Fig. 3(d)], the longitudinal compliance along $c$-axis of the crystal is decreasing up to about $\theta = 55°$ and increases further with $\theta$ marginally. The experimental measurements of $s_{33}$ in Ref. 29 also shows up similar trends of showing a maximum at $\theta = 0$ and a minimum at $\theta = 50°$ and increases further with $\theta$ as is seen from the (010) cross-section Fig. 4(c). The shear components of the compliance $s_{44}$ [Fig. 4(a)] shows greater anisotropy while rotating the $c$-axis (variation of $\theta$) compared to that during the rotation of $ab$-plane (variation of $\phi$). $s_{66}$ [Fig. 4(b)] exhibits strong anisotropy in its values when the ($\phi$, $\theta$) pair is less than about 50°. At $\theta = 90°$, $s_{66}$ has no dependence on $\phi$. Also, as a check to the evolution shown by $s_{\mu\nu}$ against the variation of Euler angles, we have plotted the stiffness $C_{\mu\nu}$ (not shown) and found to be exactly complementary to $s_{\mu\nu}$. Figure 4(d) reveals that the thickness mode electromechanical coupling factor $k_t = e_{33}/\sqrt{C_{33}^D \kappa_{33}^\varepsilon}$, where $C_{33}^D = C_{33}^E + (e_{33})^2/\kappa_{33}^\varepsilon$, is maximum in single crystals of BaTiO$_3$ while along polar direction and the rotation of *ab*-pane has hardly any impact on the coupling. It is interesting to note that in applications like transducers where the coupling is critical to the efficiency, the single crystal spontaneous polarization direction delivers the maximum energy conversion rather than a canted crystal. However, rotation of *c*-axis decreases coupling and finally put it to nil at $\theta = 90°$. Hence altogether, effects like spontaneous polarization ($\boldsymbol{P}_s$) rotation achieved through crystal orientation in the presence of an external electric field can make rapid increases in net polarization (Fig. 1) in certain directions in single crystal ferroelectrics. (Here it may be noted that the polarization-vector is dictated by $P_i = d_{i\mu}T_\mu$ in the direct piezoelectricity). However, the rotation ($\phi$) of the *ab*-plane has hardly any impact on either piezoelectric (Fig. 1) or dielectric coefficients (Fig. 2) but on compliance $s_{\mu\nu}$ (Figs. 3 and 4).

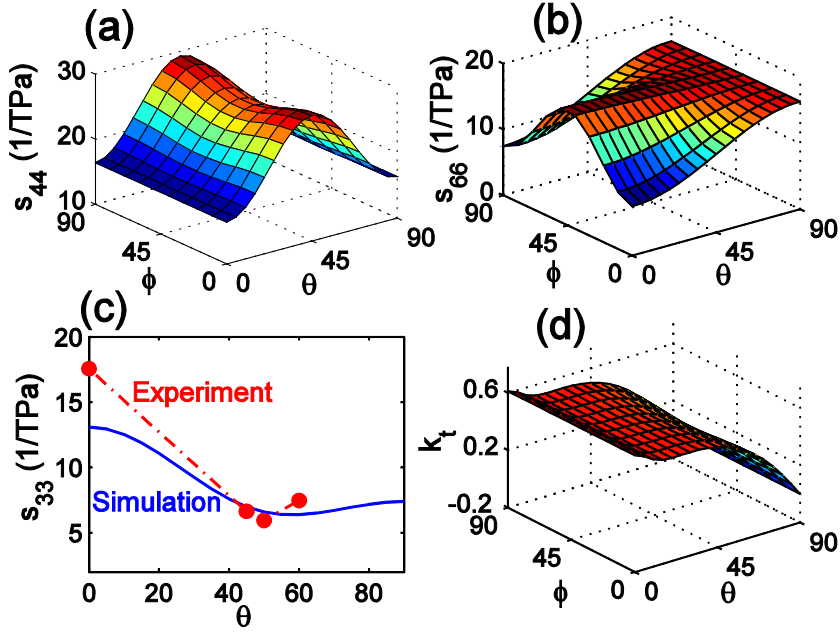

FIG.4. The variation of the elastic compliances (a) $s_{44}$, (b) $s_{66}$ with the crystal orientation angles $\phi$, $\theta$ in BaTiO$_3$ single crystal. (c) The cross section curve obtained when a cut by the crystallographic plane (010) is done to enable comparison of simulated $s_{33}$ with experiments in Ref. [29] (filled circles). (d) Electromechanical coupling factor $k_t$, variation with ($\phi$, $\theta$) in BaTiO$_3$ single crystal. Here the angles $\phi$ and $\theta$ are in degrees.

**B. Polarization rotation and the polycrystal**

Based on the good agreement of results with the published data in single crystals we proceed here to apply the methodology in ferroelectric polycrystals. Here we first discuss how the ceramic ferroelectric is canted. The fully poled polycrystal has almost all the net polarizations $\bar{\boldsymbol{P}}$s of the crystallites aligned along the z-direction (here, we should assume that polarization direction in each domain would adopt configuration closest to that of the applied field).[22] Crystallographic texturing of polycrystalline materials enables preferential orientation of the crystallites in ferroelectric ceramics.[33] In other words, the *c*-axes of the crystallites can be oriented along a desirable direction by this method. Since the spontaneous polarization-vector is lying along the *c*-axis direction of the crystal, texturing can increase the poling efficiency by enabling reorientation of polar vectors of the unaligned crystallites. Yet constraints such as grain size effects, orientation effects, intergranular interactions and the effects of inhomogeneities create local fields that can affect macroscopic piezoelectricity of the ferroelectric ceramics and impedes the poling.[22, 34] The angle $\theta$ s characterizing the orientation of $\bar{P}$s of almost all the crystallites in a fully aligned ceramic is zero. However, the *ab*-planes still obey randomness. Hence, the fully aligned polycrystal has Euler angles ($\phi, \psi$) obey a near uniform random distribution with the third Euler angle $\theta \approx 0$. However, in fact the alignment of the $\bar{P}$s will not be complete as we have envisaged as it is very difficult for the polarization to percolate through the ceramic on poling as some grains will not have possible domain states that are nearly aligned with the field [22] and hence $\theta$ will obey a normal distribution however, with a small standard deviation.[16] This is partly due to the hysteresis of ferroelectrics revealing that the zero-field remnant polarization $P_r$ is slightly less than the spontaneous polarization $P_s$. In our simulation of the ferroelectric polycrystal BaTiO$_3$ we kept the set of angles ($\phi, \psi$) at random between $-\pi$ to $+\pi$, keeping in mind the laws of coordinate transformation given in Ref. 25 and varies $\theta$ first from 0 to 90°. This is to simulate the hypothetical case of a ceramic whose crystallites are completely aligned with the poling electric field. Next step is letting $\theta$ to be a normal distribution with standard deviation $\sigma$ varying from 0.01 to 15 with worsening texture and with mean *m* varying from 0 to 90°. The progressive shifts of the mean *m* from zero will ensure the canting of the polycrystal from 0 to 90°. For instance, for a mean *m* = 15°, the canting of the respective crystallites would turn 15° (in the counter-clockwise direction) more from their respective orientations at $m = 0$ from the reference $z$-axis ([001] direction). Thus, if $m$ assumes a higher value than zero say, $m = m'$ we can certainly say that the polycrystal is canted by angle $m'$ with reference to [001]. Thus the value of the average $m$ is the measure of the orientation of the polycrystal. [27]

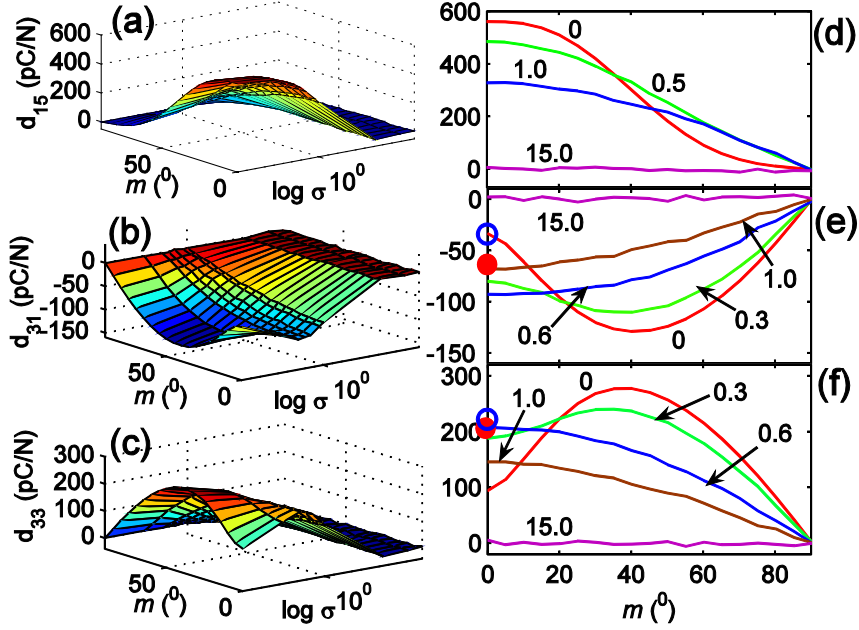

FIG. 5. (a)-(c); Variation of the homogenized piezoelectric strain coefficients $d_{jv}$ with the orientation distribution function parameters $\sigma$ and $m$ for the polycrystalline $BaTiO_3$. The subplots (d), (e) and (f) show dependence on $m$ for selected $\sigma$ (numbers written close to the plots) of $d_{15}$, $d_{31}$ and $d_{33}$ respectively. Filled circles in Figs. (e) and (f) are experimental results from Refs. [4] and [12] respectively for [111] poled $BaTiO_3$ single crystal and open circles are the peak values obtained by us for single crystals.

The dependence of the electromechanical properties on the distribution of the grain orientations of the polycrystalline $BaTiO_3$ is displayed in Figs. 5, 6 and 7 respectively. The results marked with $\sigma = 0$ corresponds to the first case where $\phi$ and $\psi$ are at random and $\theta$ varies in steps from 0 to 90°. All the three piezoelectric moduli approaches zero as the texture parameter $\sigma \to \infty$. In other words, as distribution becomes flatter and hence the orientations of net polarization of crystallites $\bar{P}$s approach fully random the aggregate piezoelectricity vanishes. Moreover, as the $\sigma \to 0$ the ferroelectric material approaches full texture and piezoelectric moduli approach single crystalline values. The variation of the mean $m$ of the orientation distribution also affects the values significantly.

With the $\sigma$ value increases we enter the realm of random polycrystals. The piezoelectric strains $d_{i\mu}$ of ceramic $BaTiO_3$ which are expected to fall in between single crystal values and zero (though it holds true for coefficients $d_{15}$) with worsening textures exhibit peaks as shown in Figs. 5. The shear piezoelectric coefficient $d_{15}$ shown in Figs. 5(a) and (d) is found to fall monotonously with both the parameters of distribution. Fig. 5(d) redraws the dependence of rotation (mean $m$ is the measure of it) of the polycrystal $d_{15}$ at selected standard deviations $\sigma$ of the orientation distribution. Here it may be recalled that it shows similar decay with the rotation $\theta$ of the polarization in single crystal $BaTiO_3$ [Fig.1 (a)]. – $d_{31}$ also show a similar decrease with $\theta$ in single crystal [Fig.1 (b)]. However, – $d_{31}$ shows peaks at the increase of both $\sigma$ and $m$ in ceramic $BaTiO_3$ as is shown in Fig. 5(b). Figure 5(e) demonstrates that there would be a host of parameter values ($\sigma$ and $m$) at which the polycrystalline ceramic $BaTiO_3$ is superior to [111] poled single crystal. All $\sigma$ values up to 1.0 can deliver higher piezoelectric coefficients – $d_{31}$ than [111] oriented single crystal value (= -62 pC/N) from Ref. 4, however at different ranges of $m$ values. The peak value of $d_{31}$ obtained is -131.8 pC/N at $m = 40°$. The longitudinal piezoelectric coefficient $d_{33}$, attains a value of 223.4 pC/N close to the [111] poled single crystal value from our simulation (= 223.7 pC/N) and from the experiment (= 203 pC/N) [12] at a mean $m = 25°$ between standard deviations $\sigma = 0.1$ and 0.3 as shown in Figs. 5(c) and (f). As shown in Fig. 5(f) $d_{33}$ keep on exhibiting higher values than [111] poled single crystal BaTiO$_3$ in the range of standard deviation $\sigma \leq 0.6$ however at different ranges (between 0 and 55°) of polycrystal rotation *m*. At $m = 40°$, $d_{33}$ attains the peak value of 275.9 pC/N. Hence we have a set of distribution parameters at which it is possible to design ferroelectric ceramics which exceeds in piezoelectricity compared to their single crystal version. This

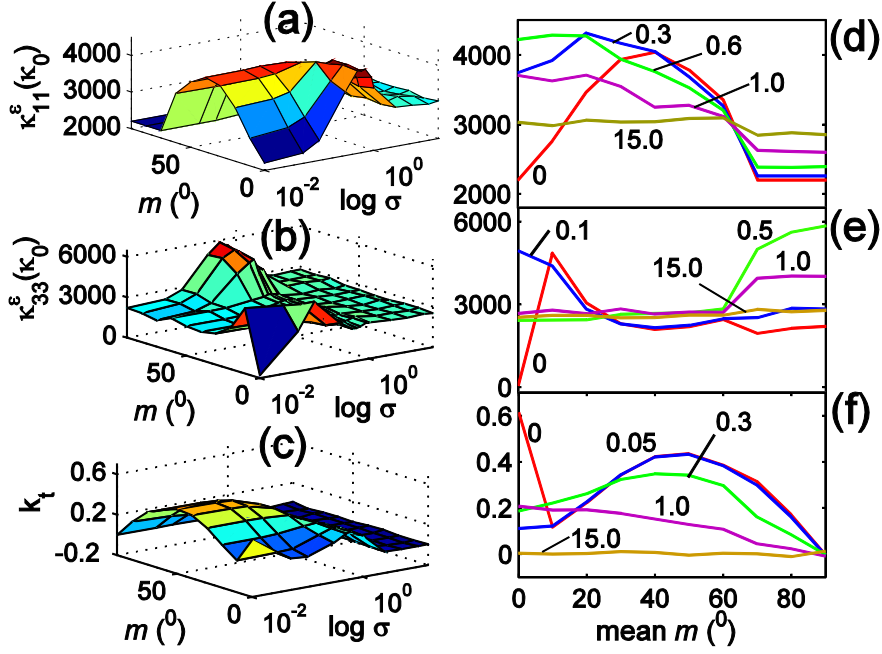

FIG. 6. Variation of the homogenized relative dielectric coefficients (a) $\kappa_{11}^{\varepsilon}(\kappa_0)$, (b) $\kappa_{33}^{\varepsilon}(\kappa_0)$ and (c) the electromechanical coupling coefficient $k_t$ with the orientation distribution function parameters $\sigma$ and *m* for the polycrystalline BaTiO$_3$. Subplots (d), (e) and (f) show the dependence on *m* of physical properties at selected $\sigma$ (numbers written close to the plots) given in Figs. 6(a)-(c) respectively.

suggests that piezoelectric properties can be optimized by a proper choice of the parameters which control the distribution of grain orientations. Nevertheless, it is impossible to analyze all possible combinations of the distribution parameters or the angles themselves. Hence we have implemented the stochastic optimization technique of simulated annealing for the optimization problem in search of a better $d_{33}$ than 275.9 pC/N. However, the optimization procedure has not yield any values better than the present one. This bears significance as $d_{33}$ is a very important piezoelectric coefficient among $d_{jv}$, determining the level of induced strain at a given electric field and a widely used figure of merit describing actuator performance. In technological applications like MEMS most of the piezoelectric thin films are polycrystalline, a controlled selection of texture can provide better piezoelectric response. In such devices, maximising the piezoelectric coefficient is of considerable importance in reducing the drive voltage or increasing the sensitivity. [35, 36]

The impact of anomalous increases in dielectric permittivity shown in Fig. 6 has a direct bearing on the piezoelectricity of ceramic $BaTiO_3$. A higher dielectric susceptibility is thought to be caused by the flattening of the elastic free energy profile and consequently results in high piezoelectric coefficients. [32, 37] Here the large increase (reaches up to 4391 from 2200 due to rotation of the polycrystal) of relative permittivity $\kappa_{11}^{\varepsilon}$ [Figs. 6(a) and (d)] perpendicular to the polar direction dominates the piezoelectric behaviour of ceramic $BaTiO_3$. It may further be noted that $\kappa_{11}^{\varepsilon}$ is invariant under the rotation in single crystal phase. Also the electromechanical coupling $k_t$ exhibits enhancement on rotation [Figs. 6(c) and (f)] from virgin polycrystal values although the value is still smaller than polar single crystal $BaTiO_3$.

The rigorous coordinate transformations in elastic compliance is dictated by $s_{ijkl} = a_{im} a_{jn} a_{kp} a_{lq} s_{mnpq}^{c}$, where $a_{jk}$ are Euler transformation tensor components [25] and $s_{mnpq}^{c}$ are the compliance tensors of crystallite along [001]. The randomness we set to the orientations of the *ab*-plane of the crystallites renders the compliance $s_{11}$ along that plane different from single crystal $BaTiO_3$ [see Fig.3 (a)]. As demonstrated in Fig. 7, $s_{11}$ shows strong dependence with mean in the case of $\sigma = 0$, which corresponds to stepwise increase of $\theta$ while keeping $\phi$ and $\psi$ at fully random. This finding is in contrast to our simulation results showing invariance of $s_{11}$ on polycrystal keeping both $\phi$ and $\psi$ at zero while varying the $\sigma$ and $m$ of $\theta$. Interestingly, other components of $s_{\mu\nu}$ behave similarly in both cases recording almost the same pattern of anisotropy with both $\sigma$ and $m$. After a certain value of $\sigma$ the compliance has hardly any dependence on either $\sigma$ or $m$.

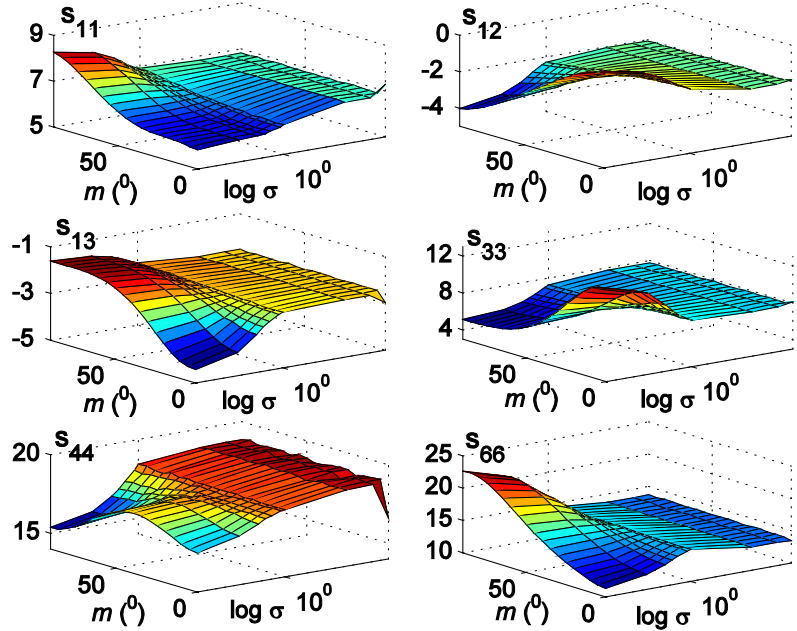

FIG. 7. Variation of the homogenized elastic compliance coefficients $s_{\mu\nu}$ [in $(TPa)^{-1}$] with the orientation distribution function parameters $\sigma$ and $m$ for the polycrystalline $BaTiO_3$.

In conclusion, we have demonstrated a design strategy utilizing the randomness judiciously to simulate ceramic ferroelectrics which are superior to optimally poled single crystals. Noncollinear rotation of polarization realized by canting of the poled polycrystal holds the key to the enhancement of physical properties. An extension of this method to relaxor FE materials with ultrahigh piezoelectricity might provide the avenue for a host of high performance ceramic piezoelectrics.[38] Our results provide guidance in tuning the orientations at the granular level to manufacture superior ceramic ferroelectrics in the backdrop of the physical property control achieved by tailoring the microstructure.

**ACKNOWLEDGMENTS**

We thank R.E. Cohen for fruitful discussions. This work was supported by FCT, Portugal under POCI 2010 and K.P.J. was supported under SFRH/BPD/20226/2004.

**APPENDIX**

The piezoelectric material functions $C_{\mu\nu}^{E}$, $e_{k\mu}$ and $\kappa_{ik}^{\varepsilon}$ be Y-periodic functions in the unit-cell defined by $Y = [0, Y_1] \times [0, Y_2] \times [0, Y_3]$ in the space $\mathbb{R}^3$ of coordinates $y_i$.[21, 24] If $\boldsymbol{y} = \boldsymbol{x}/\lambda$ where $\lambda > 0$ is a real positive parameter representing the microstructure scale, in order to find the effective properties, the piezoelectric displacement $\boldsymbol{u}^\lambda$ and electric potential $\phi^\lambda$ are expanded asymptotically, giving: $\boldsymbol{u}^\lambda = \boldsymbol{u}_0(\boldsymbol{x}) + \lambda \boldsymbol{u}_1(\boldsymbol{x}, \boldsymbol{y})$ and $\varphi^\lambda = \varphi_0(\boldsymbol{x}) + \lambda \varphi_1(\boldsymbol{x}, \boldsymbol{y})$ where only the first-order variation terms are considered and $\boldsymbol{u}_1(\boldsymbol{x}, \boldsymbol{y})$ and $\varphi_1(\boldsymbol{x}, \boldsymbol{y})$ are Y-periodic. The functions $\boldsymbol{u}_0$ and $\varphi_0$ are constant with respect to $\boldsymbol{y}$. Due to the linearity of the problem and assuming separation of variables we obtain

$$\boldsymbol{u}_1(\boldsymbol{x}, \boldsymbol{y}) = \boldsymbol{\chi}(\boldsymbol{x}, \boldsymbol{y}) \varepsilon(\boldsymbol{u}_0(\boldsymbol{x})) + \boldsymbol{\Phi}(\boldsymbol{x}, \boldsymbol{y}) \nabla_x \varphi_0(\boldsymbol{x})$$ and

$\varphi_1(\boldsymbol{x}, \boldsymbol{y}) = \psi(\boldsymbol{x}, \boldsymbol{y}) \varepsilon(\boldsymbol{u}_0(\boldsymbol{x})) + R(\boldsymbol{x}, \boldsymbol{y}) \nabla_x \varphi_0(\boldsymbol{x})$, where $\boldsymbol{\chi}(\boldsymbol{x}, \boldsymbol{y})$ is a characteristic displacement, $R(\boldsymbol{x}, \boldsymbol{y})$ is a characteristic electric potential, $\boldsymbol{\Phi}(\boldsymbol{x}, \boldsymbol{y})$ and $\psi(\boldsymbol{x}, \boldsymbol{y})$ are characteristic coupled functions of the unit-cell and $\nabla_x \varphi_0(x) \equiv \dfrac{\partial \varphi_0(x)}{\partial x_k}$, $k = 1,2,3$. These functions are solutions of the local problems derived using variational principles applied in linear piezoelectric medium and subsequent application of asymptotic analysis utilising the space derivatives of the functions $u_1(x,y)$ and $\varphi_1(x,y)$. There would be four equations as there is as much number of unknown characteristic functions. The microscopic system of equations characterizing $\chi(x,y)$, $R(x,y)$, $\Phi(x,y)$ and $\psi(x,y)$ is solved computationally using finite element method. The space derivatives of the characteristic functions enable the evaluation of macroscopic electromechanical given in Eqs. (1)- (3). As the 3D unit-cell is expected to capture the response of the entire piezoelectric system (where the macroscopic piezoelectric material is constructed by adding, contiguously to the representative unit-cell, a large number of other identical unit-cells), particular care is taken to ensure that the deformation across the boundaries of the representative unit cell are compatible with the deformation of adjacent unit-cells.